# Issues in Strategic Decision Modelling

Paula Jennings
BDO Stoy Hayward
8 Baker Street
LONDON
W1U 3LL

**ABSTRACT**

*Models are invaluable tools for strategic planning. Models help key decision makers develop a shared conceptual understanding of complex decisions, identify sensitivity factors and test management scenarios. Different modelling approaches are specialist areas in themselves. Model development can be onerous, expensive, time consuming, and often bewildering. It is also an iterative process where the true magnitude of the effort, time and data required is often not fully understood until well into the process. This paper explores the traditional approaches to strategic planning modelling commonly used in organisations and considers the application of a real-options approach to match and benefit from the increasing uncertainty in today's rapidly changing world.*

## 1    Introduction

In recent years the spreadsheet has become a familiar tool used from Boardroom to back office; in very few of those organisations has it taken the leap forward to provide the insight needed to help leaders maximise the benefits of exercising strategic choice.

With rapid technological change and global instability, strategic planning needs to support the ever-changing vision of a company. In the past a company could map it's course from A to B, and steer the ship on that path. By applying techniques and recognising some of the human limitations surrounding decision making, companies can benefit from this increased uncertainty.

The objective of modelling in the strategic planning process is to support the decision maker by providing a quantitative analysis of the strategic options and seek the optimal decision in terms of the individuals or companies objectives.

In order to achieve this objective a combination of human and technical resources are needed. As Banks and Monday (2001) highlighted the spreadsheet '….still suffers from significant defects in the way that they inform or misinform decision makers.

Strategic planning modelling has advanced from the simple model with limited ability to perform what-if analysis, to inclusion of techniques such as Monte Carlo simulation, to the application of a real options pricing approach.

## 2    Traditional spreadsheet models

The development and use of spreadsheets for strategic decision-making has long been common practice amongst finance professionals. A typical feature is the use of point estimates, the most likely values, as input assumptions.

Management, supplying these estimates, often have differing and at times diametric views on these values. Often management are expressing their personal 'best guess' even though they believe that the true value is likely to lie in a range of values, rather than a specific point. As Panko and Haverson (1996) note 'developers may even include deliberately incorrect data, or at least data from estimates that are dubious but support their case'.

In considering the need to recognise possible bias developers may wish to extend their risk analysis beyond traditional approaches to include sensitivity testing and scenario or what if analysis, techniques such as Monte Carlo simulation and Real Options approaches.

## 3    Sensitivity testing

One of the most commonly used techniques in risk analysis is sensitivity testing. The aim of this technique is to select individual assumptions and assess the outcomes sensitivity to changes in that specific assumption.

In terms of a spreadsheet model, input assumptions considered to be high risk can be highlighted. Input values can then be singularly changed and a new outcome recorded.

Sensitivity testing can be useful when trying to identify specific factors that require management focus in order to achieve a successful outcome.

A key limitation of sensitivity testing however is that of processing single input value changes. In reality a change in a singular variable alone without impact on other key inputs or indeed the market as a whole would be unusual.

Without applying a common sense approach to sensitivity testing it can become little more than a number crunching exercise to satisfy stakeholders that risk has been assessed.

In most cases the true value of such analysis is limited to projects where there is little flexibility in how risk factors can be managed and when the project will go ahead regardless of the degree of undesirable outcomes identified.

## 4    Scenario or 'what-if?' analysis

Scenario analysis moves a step further to consider clusters of risks reflected in changes in a range of assumptions. Commonly management build a picture of best case, worst case and most likely values for each input value. The model will then calculate the outcome of each scenario.

Although an improvement on single point estimates, the process of calculating every possible combination from the range of input variables can be very time consuming and result in lots of data. Furthermore it gives you no sense of the probability of achieving different outcomes.

## 5    Monte Carlo simulation

To undertake a timelier and ultimately more complete risk analysis a combination of a spreadsheet modelling and simulation may be required.

For each uncertain variable (an input value within a range of possible values) the possibly values with a probability distribution are defined. The type of distribution selected is based on the conditions surrounding the variable.

The simulation calculates multiple scenarios of a model by sampling values from the probability distribution and using the values in that cell.

By defining key output measure, known as the forecast values, the impact of the simulation on that forecast value and assessment of the certainty that a particular forecast value will fall within a certain range can be reviewed.

Application of this approach can improve sensitivity analysis, for example to show you which factors are most responsible for the uncertainty surrounding cash flows. Monte Carlo simulation allows management to gain a richer understanding of risk without the need to spend time running and analysing multiple scenarios.

Tools such as Crystal Ball alleviate the need to devise the equation that represents this distribution, making the process considerably less onerous. However simulation involves the application of statistical techniques and should be supported by a developer with the appropriate knowledge.

## 6    Optimisation

Optimisation is about identifying the best solution to the problem using the model. It is particularly appropriate when management have control of a range of variables impacting on the outcome.

Traditionally Excel Solver can be applied to generate optimal solutions by applying a linear equation. The limitation of this approach is that it is difficult to take account of the range of uncertainty surrounding these variables.

Building on our ability to assess the degree of uncertainty, tools such as OptQuest can help strategic decision makers identify the optimal decision. Again this can provide management with valuable knowledge on how to benefit from uncertainty, but must be supported by stakeholders and those with the requisite technical knowledge.

## 7    Real Options

### 7.1    Introduction
As a result of economic changes strategic planning can benefit from moving forward to include other approaches to strategic decision modelling, such as  real options valuation. This approach seeks to build on the concept of strategic choice and aims to provide a 'strategic map' of different routes that may be taken depending on choices selected through the journey.

## 7.2    Outline of a real-options approach to modelling

Originally used by the financial services sector the Black-Scholes valuation tool provided a valuable insight in capital markets. A real-options valuation approach has traditionally been used in oil and gas exploration and the management of large pharmaceutical portfolio where decisions of this nature are made frequently and are of significant value, however it is worth considering its wider application to decisions where a staged approach can be taken.

An example might be in the property sector where decisions to acquire, sell, retain or develop property are decisions dependent on information from the internal and external market, for example about demand in the housing or retail sectors.

The real options approach is based on the idea that there is value in waiting until more information is available; for this value to exist delay is making the decision must be possible. This is particularly important where high levels of uncertainty exist. Building on traditional approaches used in strategic planning modelling, real-options enables managers to take multi-contingencies into account, plan responses and phase actions accordingly.

By identifying the uncertainty, profiling the risk factors then building the financial model there is increased awareness of the 'framework' for the decision. The use of influence diagrams and decision-tree's during the process can simplify and communicate the framework of the decision to a broad range of stakeholders.

## 7.3    Application of the real-options approach

A 'real options' approach could provide great value to strategic decision makers. For those in strong market positions with a significant number of options taking this approach would provide closer to 'real-time' decision making through the use of real-time market information input to the strategic decision making process.

## 7.4    Limitations of a real-options approach

The most obvious difficulty when considering using a real options approach is the inherent complexity and confusion surrounding it. Understanding and application of modelling techniques requires specialist knowledge.

A real-options approach is also complex in terms of management's ability to identify, create and exercise the real-options.

There has also long been a conflict between corporate finance and strategic decision making approaches. Stewart Myers (Journal of Applied Corporate Finance, 2002) commented that he was 'seeing signs of convergence or growing agreement between the people in decision analysis and classis real options'.

## 8    Conclusion

In response to the changing environment we should consider the advantages and limitations of applying advanced modelling techniques to support strategic decision-making.

Encouraging all developers to incorporate sensitivity testing and 'what-if?' analysis into models will provide management with a more flexible approach to decision analysis and a greater understanding of risk.

The use of simulation and optimisation techniques provides further insight into maximising opportunity whilst minimising risk.

As economic and market uncertainty increases, and strategy becomes more flexible modelling using a real-options approach may be more widely applicable, and allow management to profit from uncertainty.

Finally it should be noted that human knowledge plays a key role in the development of spreadsheet models. As noted by Banks and Monday (2001), if knowledge is taken as 'dynamic human processing justifying personal belief toward the "truth" (Takeuchi and Nonaka in (Morely et al 2000) even after applying advanced modelling techniques developers must still view the output of a spreadsheet as 'justified personal belief' rather than 'truth'.

## 9    Case Study

### 9.1    Background
A typical example of selecting the right tool for the right purpose can be given in the following situation. The Finance Director of a multi-national company is facing an unpleasant and difficult decision in restructuring manufacturing operations. His goal is to identify the path of closure, relocation and construction of sites over the course of the next three years that will minimise his costs and maximise potential gains in efficiency.

He needs a 'strategic map' that will help him plot his course of action during that period.

### 9.2    Applying differing approaches

#### 9.2.1    Approach 1: Simple Spreadsheet
Initially he and his finance team, who all have good financial modelling skills, develop a spreadsheet model to identify the estimated costs and benefits of their top three options. This scenario would be typical of many companies given the level of spreadsheet skills commonly within organisations.

There is however a number of issues with the approach taken and the model produced – there is little scope to look at 'what-if' scenarios and sensitivity testing is limited due to the high level aggregation of costs and benefits. The Board are asked to make a decision based essentially on three separate views of potential outcomes.

#### 9.2.2    Approach 2: Advanced Spreadsheet Modelling

Feeling a little uncomfortable with the limited insight provided by this synopsis the Board ask a team of consultants to develop a model considering a wider range of possible scenarios, and perform sensitivity testing to rank each alternative.

This approach 'framed' the problem more realistically by involving a range of senior executives within the organisation including the CEO. A clear improvement on the myopic approach taken by the Finance team.

This approach lead to the need for greater volumes of data and generation of ranges of possible outcomes using techniques such as Monte Carlo simulation, however it still did not take account of the fact that the restructuring programme could be completed using a staged investment approach. With the high degree of uncertainly remaining, even after the gathering of vast amounts of data, the conclusions reached were still inconclusive and inadequate for use as a 'strategic map' for the future.

### 9.2.3    Option3: Spreadsheet Model – Real Options Valuation

As the decisions about restructuring did not all need to be made at once, a staged approach could have been taken. This would have allowed the model to both quantify the best outcome and the worst based on estimates and known information at the time. However, as the project progressed much of the uncertainty would be resolved. For example, in their past problems had been experienced with relocation of particular operations that resulted in a significant loss of production. Rather than using the Option2 model at a single pint in time, taking real options valuation approach could allow the CEO and his team to benefit from their learning by taking account of it as they move forward through their plan.

### 9.3    Conclusion
Given the high degree of uncertainty faced and the opportunity to make a staged decision the use of approaches 1 and 2 did not provide the insight and opportunity for gain achieved by approach 3. By rushing through to reach a conclusion taking a 'strategy by numbers' approach, the value of the journey was not truly recognised nor the active learning by the company as the restructuring plan unfolded.